\documentclass[
 reprint, 
superscriptaddress,
 amsmath,amssymb,
 aps, prd,
floatfix,nofootinbib,
]{revtex4-2}

\usepackage{amsmath,amssymb,graphicx}
\usepackage{hyperref}

\begin{document}

\title{An AI-based Detector Simulation and Reconstruction\\ Model for the ALEPH Experiment at LEP}

\author{Ya-Feng Lo}
\email{ya-feng.lo@cern.ch}
\affiliation{Department of Physics and Astronomy, University of California, Los Angeles, CA 90095, USA}
\affiliation{Department of Physics, Stanford University, Stanford, CA 94305, USA}
\affiliation{Department of Particle Physics and Astrophysics, Stanford University, Stanford, CA 94305, USA}

\author{Dmitrii Kobylianskii}
\email{dmitry.kobylyansky@weizmann.ac.il}
\affiliation{Department of Particle Physics \& Astrophysics, Weizmann Institute of Science, Rehovot, Israel}

\author{Benjamin Nachman}
\email{nachman@stanford.edu}
\affiliation{Department of Particle Physics and Astrophysics, Stanford University, Stanford, CA 94305, USA}
\affiliation{Fundamental Physics Directorate, SLAC National Accelerator Laboratory, Menlo Park, CA 94025, USA}

\author{Eilam Gross}
\affiliation{Department of Particle Physics \& Astrophysics, Weizmann Institute of Science, Rehovot, Israel}

\date{\today}

\begin{abstract}
We present the application of Parnassus, a generative model for full detector simulation and reconstruction, to the ALEPH detector at the Large Electron-Positron Collider (LEP). Training on simulated $e^+e^- \to Z \to q\bar{q}$ events processed through the ALEPH detector simulation and reconstruction, we demonstrate that Parnassus faithfully reproduces the detector response at the event, jet, and particle levels, with substantially better agreement than the Delphes fast simulation. The clean $e^+e^-$ environment---free of pileup and characterized by simple event topologies---provides a well-controlled benchmark for evaluating the generative model's fidelity. Our results demonstrate that modern neural-network-based generative simulation approaches, developed primarily for LHC experiments, generalize naturally to historical collider experiments with distinct detector geometries and physics environments. This work shows that Parnassus can be applied beyond the LHC context and serves as an important tool for legacy data analysis where archival software tools are challenging to resurrect.
\end{abstract}

\maketitle
\section{Introduction}

Full detector simulation based on GEANT4~\cite{GEANT4:2002zbu} provides high fidelity but is computationally expensive, motivating the development of machine-learning-based fast-simulation techniques~\cite{Paganini:2017hrr,Krause:2024avx,Hashemi:2023rgo}.  Reconstruction algorithms are also increasingly complex and can be comparably slow to simulation.

Particle-flow Neural Assisted Simulations (Parnassus) is a framework that emulates the detector response and reconstruction of particle physics experiments using deep generative models~\cite{Kobylianskii:2024sup,Dreyer:2024bhs,Dreyer:2025zhp}.  Previous studies have shown that Parnassus produces high-quality event samples for the CMS detector~\cite{CMS:2008xjf}, outperforming the classical fast simulation and reconstruction framework Delphes~\cite{deFavereau:2013fsa}.

A key question is whether such models generalize to fundamentally different experimental conditions. The ALEPH detector at the Large Electron-Positron Collider (LEP) provides a stringent test: the $e^+e^-$ environment is essentially pileup-free, event topologies are dominated by two-jet configurations at the $Z$ pole, and the detector geometry and response differ significantly from those at the LHC.

In this work, we build a Parnassus model of the ALEPH detector, simulated with Geant~3~\cite{Brun:1987ma} and reconstructed with an energy-flow algorithm~\cite{ALEPH:1994ayc}.  Like Delphes, Parnassus is modular and each detector is specified by a neural network model that is then interfaced with a common framework. Unlike Delphes, Parnassus models are tuned to match full detector simulation samples automatically through neural network training.  An interesting feature of the ALEPH dataset is that no public collaboration-derived fast simulation program exists and the public full simulation samples are limited in number.  An accurate fast simulation and reconstruction program would enhance the growing interest in these legacy data~\cite{Electron-PositronAlliance:2019cpi,Electron-PositronAlliance:2021kig,Electron-PositronAlliance:2023klx,Electron-PositronAlliance:2025hze,Electron-PositronAlliance:2025fhk,Badea:2026klb,Defranchis:2026wyw,Moreno:2026mqk}.  Applying Parnassus to LEP also probes the portability of the Parnassus architecture.

This paper is organized as follows.  Section~\ref{sec:data} briefly introduces the ALEPH dataset and Sec.~\ref{sec:methods} describes how it is processed for the machine learning training.  Numerical results are presented in Sec.~\ref{sec:results} and the paper ends with conclusions and outlook in Sec.~\ref{sec:conclusions}.

\begin{figure}
    \centering
    \includegraphics[width=0.95\linewidth]{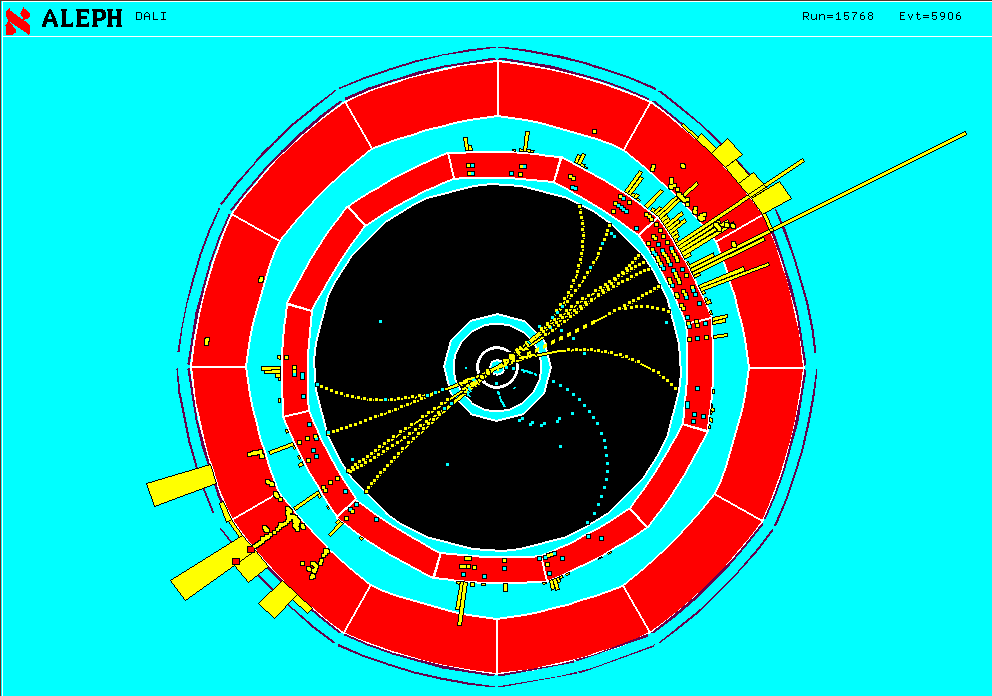}
    \caption{An event display of a collision in the ALEPH detector.  Reproduced from \url{https://aleph.web.cern.ch/aleph/aleph/newpub/dali-displays.html}.}
    \label{fig:eventdisplay}
\end{figure}

\section{Dataset and Model}
\label{sec:data}

The training dataset consists of simulated $e^+e^- \to Z \to q\bar{q}$ events at $\sqrt{s} \approx 91.2~\mathrm{GeV}$ processed through a full ALEPH detector simulation and reconstruction~\cite{ALEPH:1994ayc}.

The ALEPH detector~\cite{ALEPH:1990ceb} was one of the four experiments at LEP, operating at center-of-mass energies around the $Z$ pole. It featured a cylindrical geometry with a high-precision tracking system based on a time projection chamber, surrounded by electromagnetic and hadronic calorimeters and a superconducting solenoid providing a 1.5~T magnetic field. The detector was designed for excellent charged-particle momentum resolution and nearly hermetic energy measurements, enabling precise reconstruction of hadronic final states. A representative hadronic event is shown in Fig.~\ref{fig:eventdisplay}.

In contrast to LHC experiments, the LEP $e^+e^-$ environment is characterized by the absence of pileup and relatively simple event topologies, with hadronic $Z$ decays predominantly forming well-isolated two-jet configurations. This provides a complementary validation regime to previous LHC-based studies, enabling a particularly clean assessment of detector-response modeling in a low-occupancy environment while also posing distinct challenges due to the different detector geometry and energy scale.

Beyond serving as a clean and well-controlled benchmark, the legacy nature of the ALEPH dataset provides an additional motivation for this study, as modern fast-simulation tools for historical experiments are often limited in availability. ALEPH therefore offers an ideal testbed for evaluating whether learned detector-response models can generalize beyond the LHC and serve as practical surrogates for renewed studies of legacy datasets.

\section{Methods}
\label{sec:methods}

\subsection{Preprocessing}

The study is based on EDM4hep samples~\cite{Gaede:2022leb} of hadronic $e^+e^- \to Z \to q\bar{q}$ events processed through an ALEPH detector simulation. Each event contains reconstructed particle-flow candidates and their corresponding Monte Carlo truth information.

All particles are required to satisfy $|\eta| < 2.7$ cut, additionally we employ a $p_T > 1.0~\mathrm{GeV}$ for reconstructed particles and $p_T > 0.5~\mathrm{GeV}$ cut for truth particles. Neutrinos are removed from both collections. The remaining particles are grouped into standard categories---charged hadrons, electrons, muons, neutral hadrons, and photons---following the particle-flow convention.

After preprocessing, the dataset comprises $1{,}005{,}986$ usable events. From this pool, we define an $804{,}789$-event training sample, a $143{,}713$-event validation sample, and a $57{,}484$-event evaluation sample, ensuring that both particle- and event-level networks are trained and evaluated on statistically consistent datasets.

\subsection{Generative model}

Parnassus employs a flow-matching generative architecture~\cite{lipman2022flow, Albergo2022StochasticInterpolants} with a transformer backbone to model correlations across reconstructed objects. The framework consists of two coupled components: a particle-level conditional flow-matching network and an event-level flow-matching network. The particle-level model generates variable-length sets of reconstructed particles conditioned on global event features, while the event-level model enforces consistency of aggregate observables such as multiplicity and missing transverse momentum.

The model is trained to learn the conditional mapping from generator-level (truth) particle information to reconstructed (detector-level) observables, effectively modeling the detector response at the full-event level.

Events are represented as sets of particle features. For each particle, we record kinematic variables\footnote{These are standard coordinates for $pp$ and $ep$, but they also seem to work well in this cylindrically symmetric setting.} $(p_T, \eta, \phi)$, mass, charge, and vertex coordinates $(v_x, v_y, v_z)$. Both networks are trained using a flow-matching objective with shared normalization and conditioning variables, enabling coherent modeling of fluctuations across different levels of event description, from individual particles to global observables.

\subsection{Postprocessing}

Fast-simulation events are generated by first sampling particle-level configurations and subsequently propagating the corresponding global representation through the event-level network. The generated particles are required to satisfy $|\eta| < 3$ and are clustered into jets using the anti-$k_T$ algorithm~\cite{jetcluster2} with radius parameter $R = 0.5$.

To facilitate detailed comparisons, generated particles are matched to truth particles using a distance-based assignment within a $\Delta R$ cone. The resulting dataset contains both truth and fast-simulated collections, together with derived jet and event-level observables, enabling validation across multiple levels of granularity.

\section{Results}
\label{sec:results}

In this section, we present a detailed comparison of the ALEPH particle-flow reference simulation, the Parnassus fast-simulation output, and the Delphes fast-simulation baseline across multiple levels of granularity~\footnote{The same particle-level events are used for all three by extracting the truth from the ALEPH samples and converting to HEPMC~\cite{Dobbs:2001ck}.  The ALEPH Delphes card is from \url{https://github.com/delphes/delphes/blob/master/cards/delphes_card_ALEPH.tcl}.}. We first examine global event-level observables, followed by jet-level kinematic and substructure variables, and finally particle-level distributions and vertex information. This progression enables a systematic assessment of the model performance from coarse global properties to the most detailed detector-level observables.

The detector response is quantified using two types of residuals. For continuous observables, we use the normalized residual, which measures the fractional deviation of the reconstructed value from truth. For integer-valued observables ($N_\text{part}$, $N_\text{jet}$), we use the absolute residual.

\begin{equation}
\mathrm{Residual}(x)=
\begin{cases}
\dfrac{Y_{\mathrm{reco}}(x)-Y_{\mathrm{truth}}(x)}
{Y_{\mathrm{truth}}(x)},
&\!\!\begin{array}{l}
\text{continuous}
\end{array}
\\[0.2cm]
Y_{\mathrm{reco}}(x)-Y_{\mathrm{truth}}(x),
&\!\!\begin{array}{l}
\text{discrete}
\end{array}
\end{cases}
\end{equation}

where $Y(x)$ denotes the histogram yield in bin $x$, corresponding to the number of events, jets, or particles depending on the observable level.

\setcounter{subsection}{0}
\subsection{Event-level observables}

\begin{figure*}[t]
\centering
\includegraphics[width=0.9\textwidth]{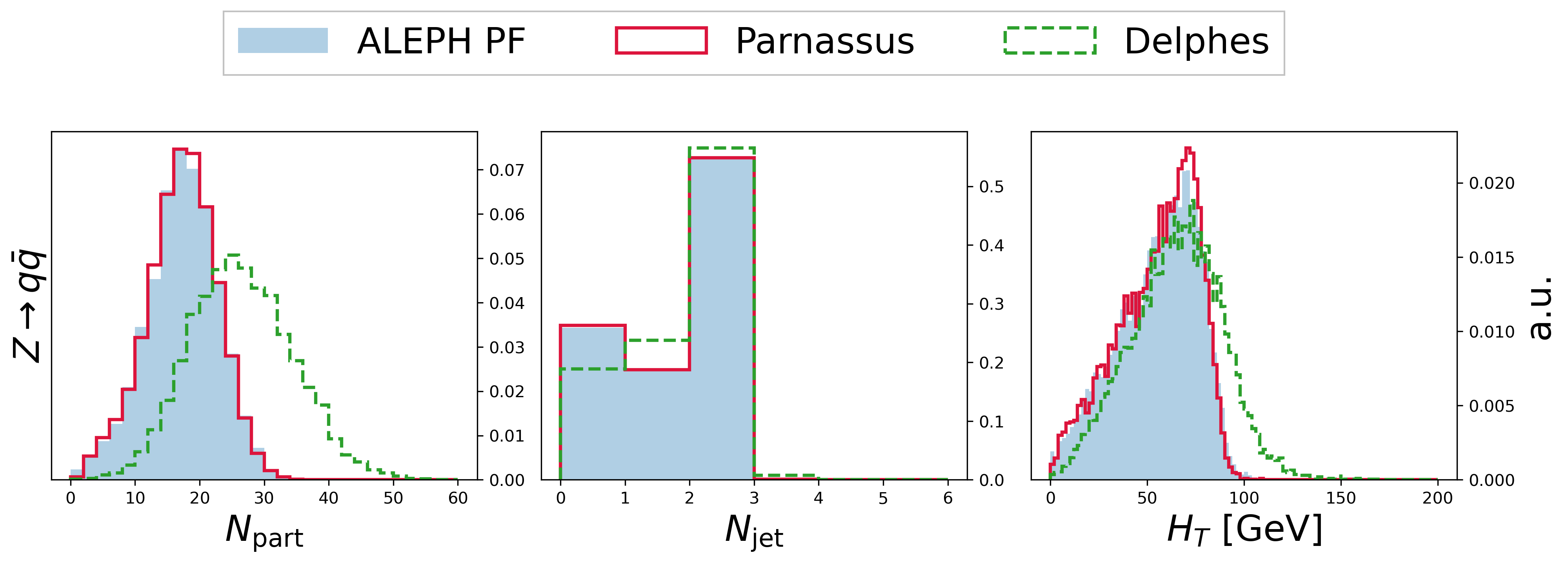}

\vspace{0.3cm}

\includegraphics[width=0.9\textwidth]{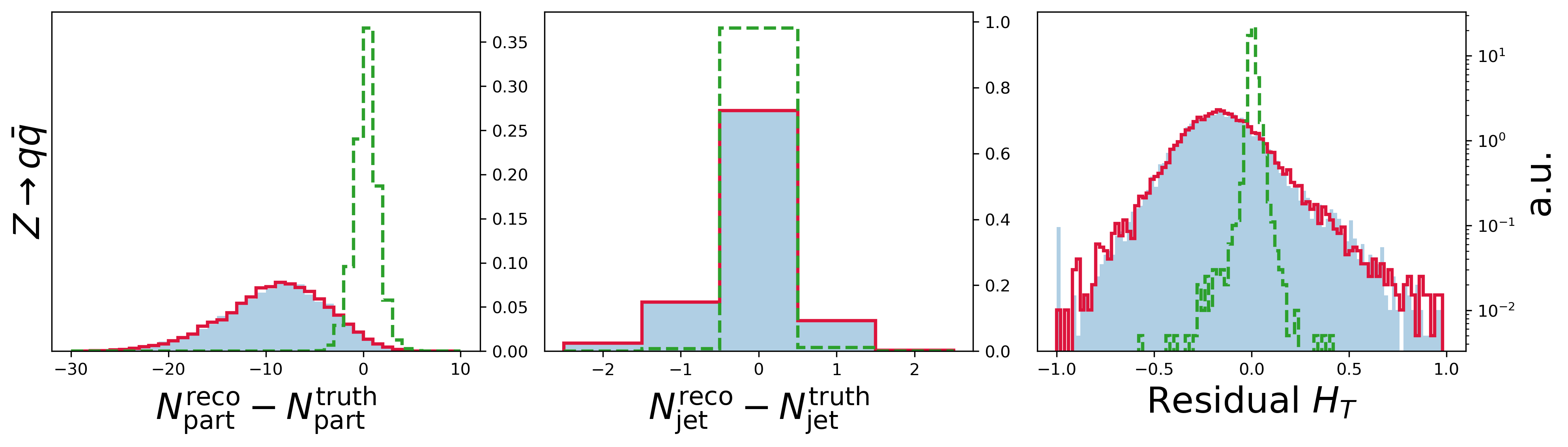}

\caption{
Event-level results. Comparison between the ALEPH particle-flow reference simulation (blue filled), Parnassus (red line), and Delphes simulation (green line). 
\textit{Top row}: Absolute distributions of $N_{\mathrm{part}}$, $N_{\mathrm{jet}}$, $H_T$.
\textit{Bottom row}: Residual distributions for $N_{\mathrm{part}}^{\mathrm{reco}} - N_{\mathrm{part}}^{\mathrm{truth}}$, $N_{\mathrm{jet}}^{\mathrm{reco}} - N_{\mathrm{jet}}^{\mathrm{truth}}$, and normalized residuals for $H_T$.
}
\label{fig:event}
\end{figure*}

\begin{figure*}[t]
\centering
\includegraphics[width=0.9\textwidth]{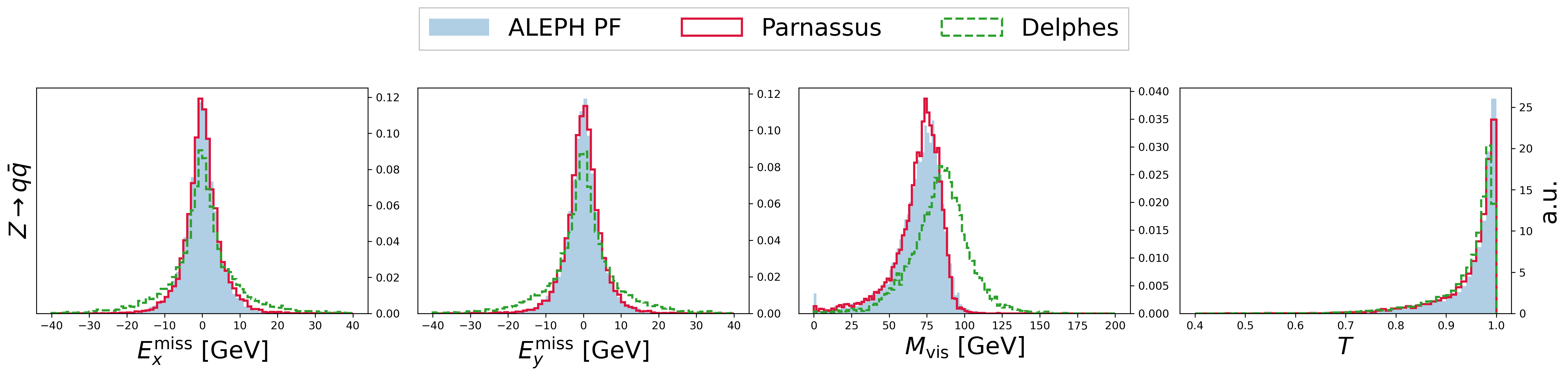}

\vspace{0.3cm}

\includegraphics[width=0.9\textwidth]{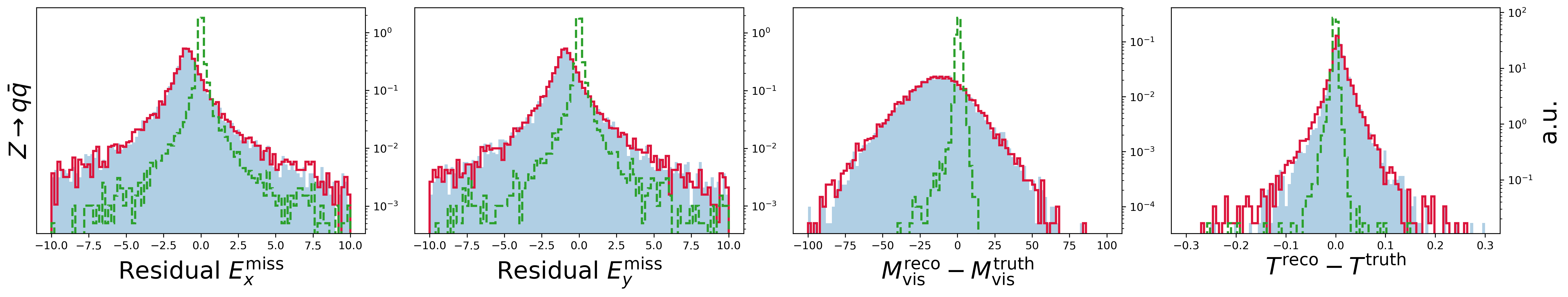}

\caption{
Event-level results. Comparison between the ALEPH particle-flow reference simulation (blue filled), Parnassus (red line), and Delphes simulation (green line). 
\textit{Top row}: Absolute distributions of $E_x^{\mathrm{miss}}$, $E_y^{\mathrm{miss}}$, $M_{\mathrm{vis}}$, and thrust $T$. 
\textit{Bottom row}: Residual distributions for $E_x^{\mathrm{miss}}$, $E_y^{\mathrm{miss}}$, $M_{\mathrm{vis}}$, and $T$.
}
\label{fig:event_thrust}
\end{figure*}

Fig.~\ref{fig:event} and Fig.~\ref{fig:event_thrust} show global observables including $N_{\mathrm{part}}$, $N_{\mathrm{jet}}$, $E_x^{\mathrm{miss}}$, $E_y^{\mathrm{miss}}$, and $H_T$. The fast-simulation sample reproduces both the shapes and residual distributions across the full kinematic range.

The particle multiplicity $N_{\mathrm{part}}$ spans a broad range, peaking at low values with a long tail, which is accurately modeled by Parnassus. The jet multiplicity $N_{\mathrm{jet}}$ is dominated by two-jet events, as expected for $Z \to q\bar{q}$ decays at the pole, with smaller contributions from higher-multiplicity configurations arising from gluon radiation; both the dominant peak and the tails are well reproduced. The missing transverse energy components $E_x^{\mathrm{miss}}$ and $E_y^{\mathrm{miss}}$ are centered near zero with symmetric tails reflecting detector resolution, and the model captures both the symmetry and width of these distributions. The scalar sum $H_T$ is well described across its full range.
The visible mass $M_{\mathrm{vis}}$ and thrust $T$~\cite{Farhi:1977sg}, which are particularly important observables for hadronic $Z$ decays at LEP, are both very well reproduced. The $M_{\mathrm{vis}}$ distribution is modeled with excellent agreement in both peak position and width. The thrust distribution, which is highly sensitive to the characteristic two-jet topology of LEP hadronic events, is also accurately reproduced, demonstrating that Parnassus captures the global event-shape structure with high fidelity.

Residuals for all event-level observables remain centered near zero, including in the low-activity regime characteristic of LEP collisions, demonstrating accurate modeling of the global event topology.

\subsection{Jet-level observables}

\begin{figure*}[t]
\centering

\includegraphics[width=0.9\textwidth]{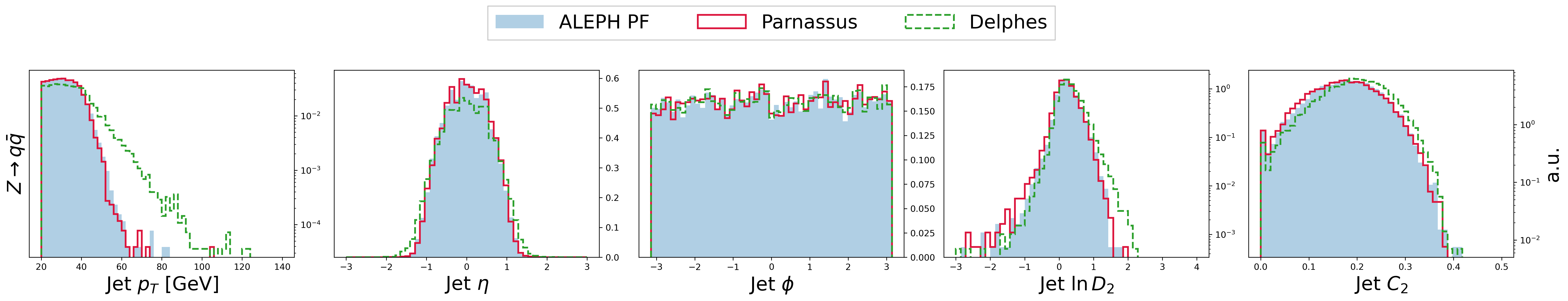}

\vspace{0.3cm}

\includegraphics[width=0.9\textwidth]{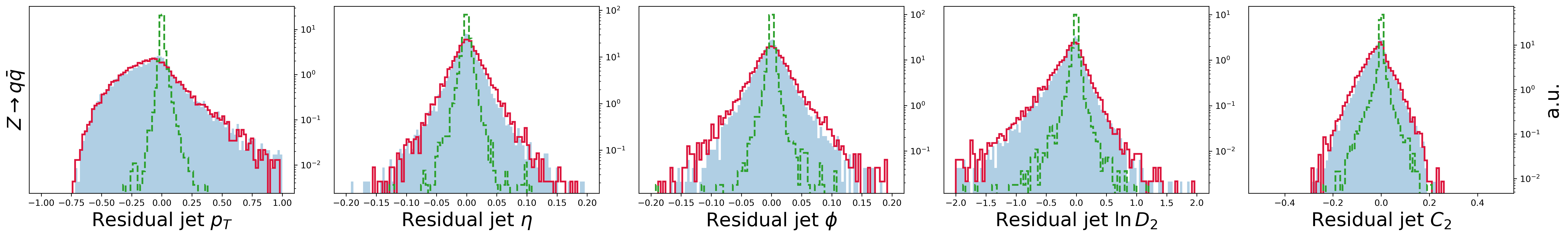}

\caption{
Jet-level results. Comparison between the reference simulation (blue filled) and Parnassus (red line), and Delphes simulation (green line). 
\textit{Top row}: Absolute distributions of jet $p_T$, $\eta$, $\phi$, $\ln D_2$, and $C_2$. 
\textit{Bottom row}: Corresponding normalized residuals.
}
\label{fig:jet}
\end{figure*}

Jet-level observables are shown in Fig.~\ref{fig:jet}. The $p_T$, $\eta$, and $\phi$ distributions are well reproduced.

The jet $p_T$ spectrum falls steeply from a peak near $\sim 25~\mathrm{GeV}$, reflecting the kinematic constraint that each jet carries approximately half the $Z$ mass. The distribution is well reproduced over roughly two orders of magnitude. The jet $\eta$ distribution is approximately Gaussian and symmetric about zero, consistent with the isotropic $e^+e^-$ collision geometry, and is accurately modeled. The jet $\phi$ distribution is uniform, as expected from azimuthal symmetry.

The jet substructure observables $\ln D_2$~\cite{Larkoski:2014gra} and $C_2$~\cite{Larkoski:2013eya} provide more discriminating tests of the model's ability to reproduce internal jet structure. The $\ln D_2$ distribution, sensitive to two-prong substructure, peaks near $\ln D_2 \sim 0$ and is well described on both linear and logarithmic scales. The $C_2$ observable, sensitive to soft-gluon radiation, peaks near $C_2 \sim 0.2$--$0.3$, and this shape is accurately reproduced.

Residuals at the jet level show slightly larger fluctuations than at the event level, particularly in kinematic tails, but remain at the few-percent level across most of the distribution, indicating good overall agreement.

\subsection{Particle-level observables}

\begin{figure*}[t]
\centering

\includegraphics[width=0.8\textwidth]{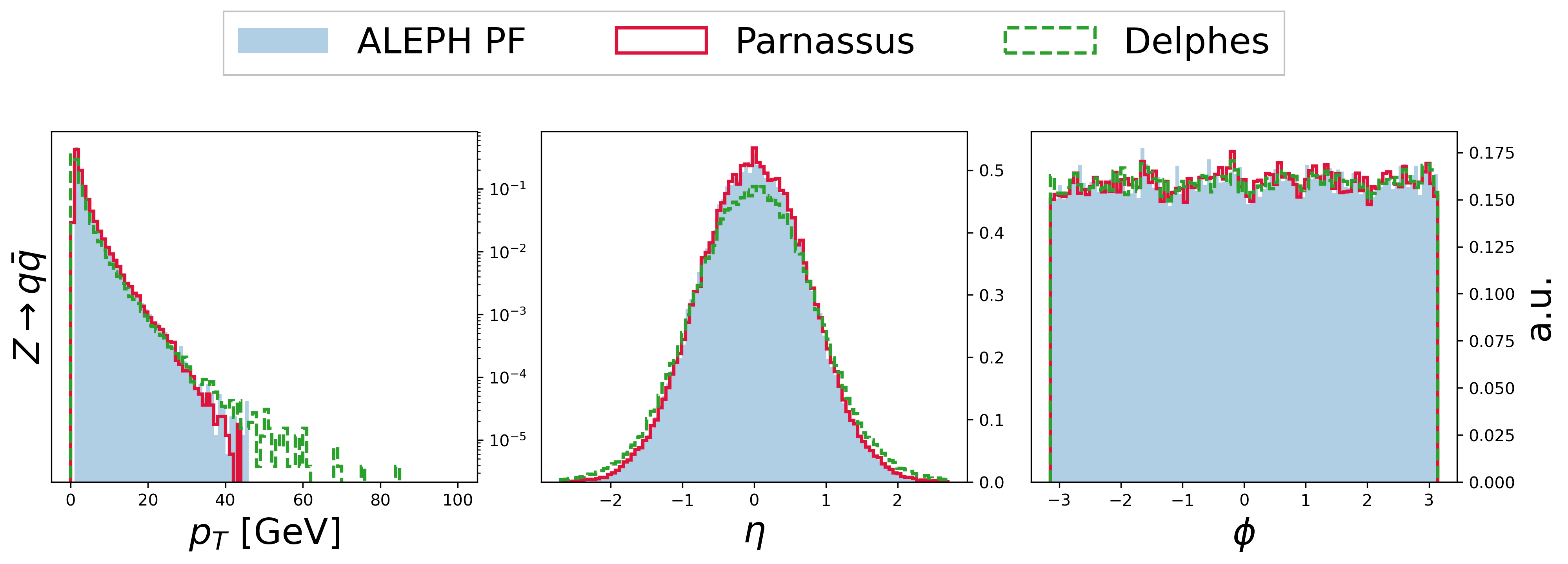}

\vspace{0.3cm}

\includegraphics[width=0.8\textwidth]{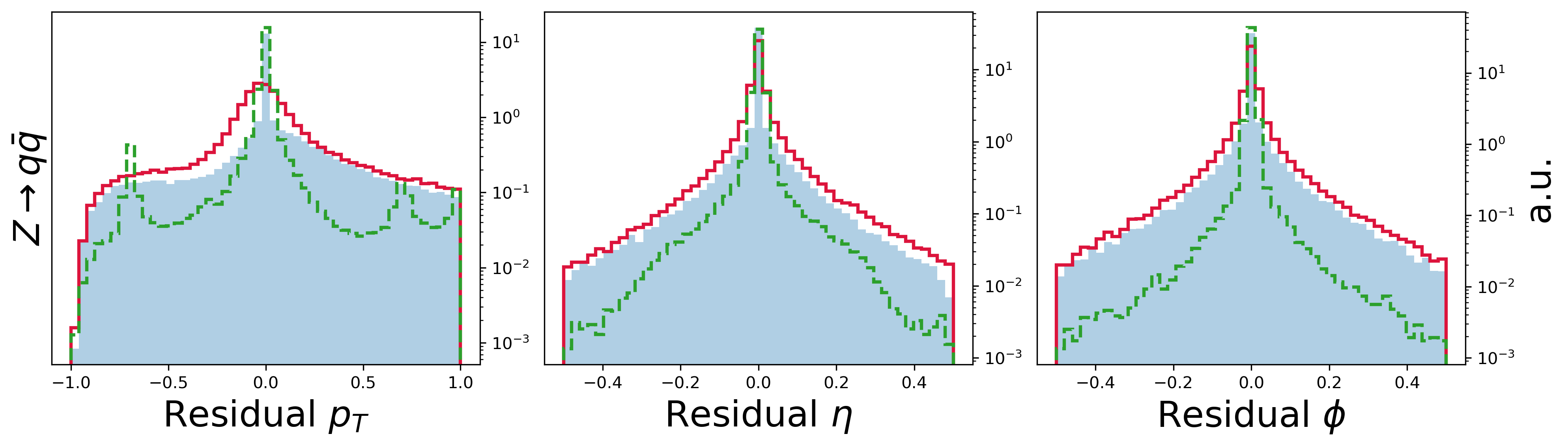}

\caption{
Particle-level results. Comparison between the reference simulation (blue filled) and Parnassus (red line), and Delphes simulation (green line). 
\textit{Top row}: Absolute distributions of particle $p_T$, $\eta$, $\phi$. 
\textit{Bottom row}: Corresponding normalized residuals.
}
\label{fig:particle}
\end{figure*}

\begin{figure*}[t]
\centering

\includegraphics[width=0.8\textwidth]{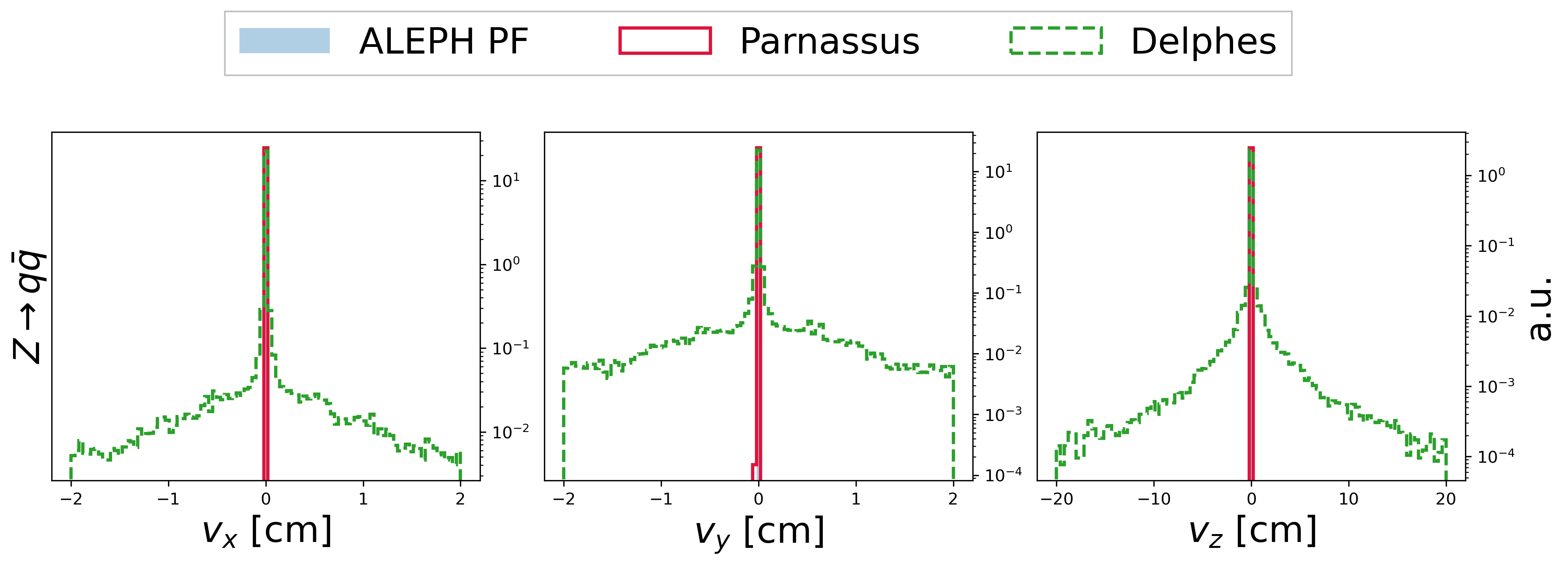}

\vspace{0.3cm}

\includegraphics[width=0.8\textwidth]{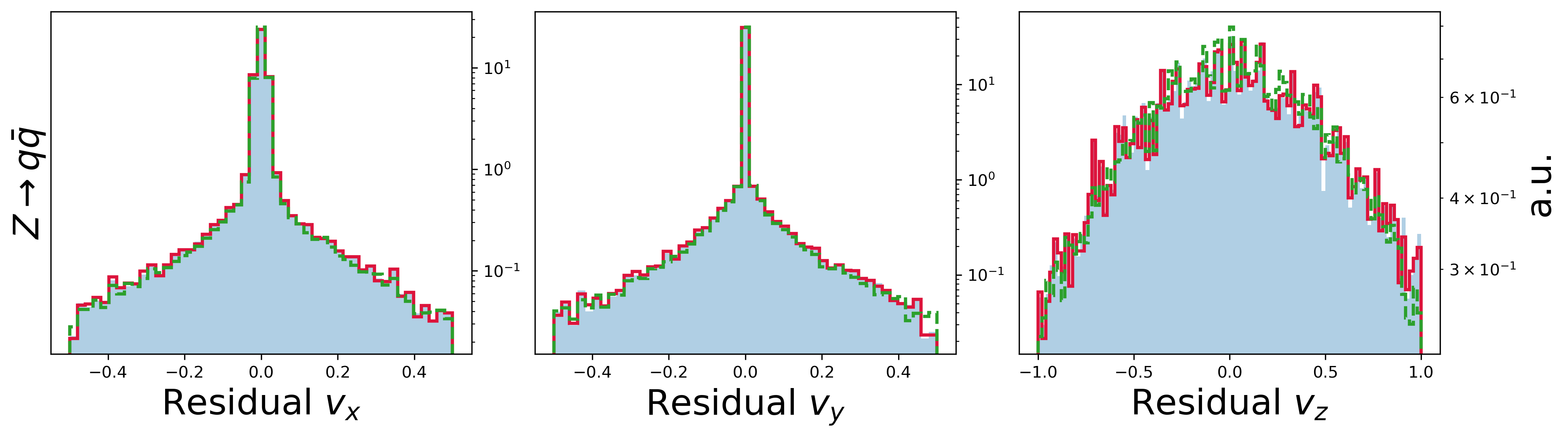}

\caption{
Particle-level results. Comparison between the reference simulation (blue filled) and Parnassus (red line), and Delphes simulation (green line). 
\textit{Top row}: Absolute distributions of particle $v_x$, $v_y$, and $v_z$. 
\textit{Bottom row}: Corresponding normalized residuals.
}
\label{fig:particle_vertex}
\end{figure*}

The most granular test is provided by individual particle-level observables, shown in Fig.~\ref{fig:particle} and Fig.~\ref{fig:particle_vertex}. Particles span a $p_T$ range from near zero to above $100~\mathrm{GeV}$, with the distribution falling over more than four orders of magnitude. Parnassus reproduces this steeply falling spectrum accurately across the full dynamic range. The particle $\eta$ distribution is peaked near zero with a broader spread than jets and is well modeled, while the particle $\phi$ distribution is flat, as expected from azimuthal symmetry. The accurate modeling of the particle-level $p_T$ spectrum over several orders of magnitude demonstrates that the learned detector response preserves both the soft and hard components of the hadronic final state. This is particularly important for LEP event-shape analyses, where soft radiation can significantly affect global observables such as thrust and jet substructure.

The vertex displacement coordinates $(v_x, v_y, v_z)$ probe the model's treatment of secondary vertices arising from long-lived particle decays. The distributions of $v_x$ and $v_y$ are strongly peaked at zero due to prompt tracks, while $v_z$ exhibits a broader distribution reflecting the extended interaction region along the beam axis. All three vertex coordinates are well reproduced, including the sharp central spike structures in $v_x$ and $v_y$. The improved agreement relative to Delphes in these vertex-sensitive observables further highlights the advantage of learned detector-response modeling for capturing fine-grained spatial information.

Particle-level residuals display broader fluctuations than at coarser granularities, particularly in distribution tails, as expected given the larger dynamic range of the observables.

\begin{figure*}[!ht]
\centering

\includegraphics[width=0.8\textwidth]{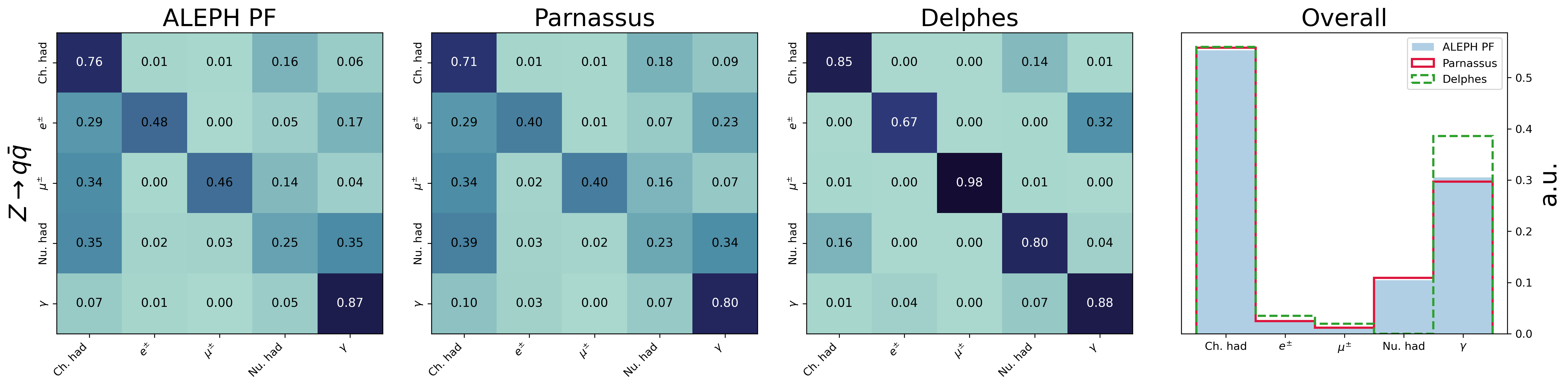}

\caption{
Confusion matrices for particle identification in $Z \to q\bar{q}$ events. Rows correspond to truth particle types and columns to reconstructed types; each row is normalized to unity such that the diagonal gives the per-class identification efficiency. Results are shown for the ALEPH PF reference simulation (left), Parnassus (center), and Delphes (right). The rightmost panel shows the overall reconstructed particle class composition for each simulation, normalized to unit area.
}
\label{fig:particle_class}
\end{figure*}

Fig.~\ref{fig:particle_class} shows the confusion matrices for particle identification, comparing the class assignment of matched truth–reconstruction pairs across the three simulations. The rows correspond to truth particle types and the columns to reconstructed types, with each row normalized to unity. Parnassus closely reproduces the per-class identification efficiencies (diagonal entries) and misidentification rates (off-diagonal entries) of the reference ALEPH PF simulation across all five particle species, whereas Delphes shows systematic deviations, indicating that its parametric detector model does not fully capture the ALEPH detector's particle identification behavior.

Overall, the Parnassus fast-simulation sample demonstrates excellent agreement with the ALEPH particle-flow reference simulation across event-, jet-, and particle-level observables, with substantially improved fidelity relative to the Delphes baseline.
\section{DISCUSSION AND CONCLUSIONS}
\label{sec:conclusions}

This study demonstrates that the Parnassus framework generalizes successfully from the LHC environment to the substantially different ALEPH detector at LEP. The model maintains high fidelity across event-, jet-, and particle-level observables despite the distinct detector geometry, lower center-of-mass energy, and characteristic two-jet topology of hadronic $Z$ decays.

The strong agreement observed for LEP-relevant observables, particularly the visible mass $M_{\mathrm{vis}}$, thrust, jet substructure, and vertex-sensitive particle-level variables, and particle identification efficiencies, highlights the ability of learned detector-response models to capture detailed experimental signatures beyond the conditions on which they were originally developed.

These results show how we can apply modern generative fast-simulation methods to legacy collider experiments and provide a practical pathway toward renewed reinterpretation and precision studies using historical datasets.


\section*{DATA AND CODE AVAILABILITY}

The code used for data preprocessing, model training, inference, and figure production is available at \url{https://github.com/parnassus-hep/cms-flow-evt}. 

\section*{ACKNOWLEDGMENTS}

We thank the ALEPH collaboration for providing the simulation datasets.  We additionally thank Sitian Qian, Sanghyun Ko, and Jingyu Zhang for useful discussions.  This work was supported in part by the U.S. Department of Energy (DOE) under Contract No. DE-AC02-76SF00515 and Award No. DE-SC0009937, and by the U.S. National Science Foundation under Grant No. OAC-2417682. DK is supported by the Minerva Grant 715027, and the Knell Family Institute for Artificial Intelligence. This research used resources of the National Energy Research Scientific Computing Center (NERSC), a DOE Office of Science User Facility supported by the Office of Science of the U.S. Department of Energy under Contract No. DE-AC02-05CH11231, under NERSC award HEP-ERCAP0035546.

\clearpage
\bibliographystyle{cms_unsrt}
\bibliography{references}

@article{ALEPH:1990ceb,
    author = "Decamp, D. and others",
    collaboration = "ALEPH",
    title = "{ALEPH: A detector for electron-positron annihilations at LEP}",
    doi = "10.1016/0168-9002(90)91831-U",
    journal = "Nucl. Instrum. Meth. A",
    volume = "294",
    pages = "121--178",
    year = "1990"
}

@article{jetcluster2,
   title="{The anti-$k_t$ jet clustering algorithm}",
   volume={2008},
   ISSN={1029-8479},
   url={http://dx.doi.org/10.1088/1126-6708/2008/04/063},
   DOI={10.1088/1126-6708/2008/04/063},
   number={04},
   journal={Journal of High Energy Physics},
   publisher={Springer Science and Business Media LLC},
   author={Cacciari, Matteo and Salam, Gavin P and Soyez, Gregory},
   year={2008},
   month=apr, pages="063--063" }

@article{Farhi:1977sg,
    author = "Farhi, Edward",
    title = "{A QCD Test for Jets}",
    reportNumber = "HUTP-77-A059",
    doi = "10.1103/PhysRevLett.39.1587",
    journal = "Phys. Rev. Lett.",
    volume = "39",
    pages = "1587--1588",
    year = "1977"
}

@article{Larkoski:2013eya,
    author = "Larkoski, Andrew J. and Salam, Gavin P. and Thaler, Jesse",
    title = "{Energy Correlation Functions for Jet Substructure}",
    eprint = "1305.0007",
    archivePrefix = "arXiv",
    primaryClass = "hep-ph",
    reportNumber = "MIT-CTP-4446, CERN-PH-TH-2013-066, LPN13-026",
    doi = "10.1007/JHEP06(2013)108",
    journal = "JHEP",
    volume = "06",
    pages = "108",
    year = "2013"
}

@article{Larkoski:2014gra,
    author = "Larkoski, Andrew J. and Moult, Ian and Neill, Duff",
    title = "{Power Counting to Better Jet Observables}",
    eprint = "1409.6298",
    archivePrefix = "arXiv",
    primaryClass = "hep-ph",
    reportNumber = "MIT--CTP-4588",
    doi = "10.1007/JHEP12(2014)009",
    journal = "JHEP",
    volume = "12",
    pages = "009",
    year = "2014"
}

@article{Dobbs:2001ck,
    author = "Dobbs, M. and Hansen, J. B.",
    title = "{The HepMC C++ Monte Carlo event record for High Energy Physics}",
    reportNumber = "ATL-SOFT-2000-001",
    doi = "10.1016/S0010-4655(00)00189-2",
    journal = "Comput. Phys. Commun.",
    volume = "134",
    pages = "41--46",
    year = "2001"
}

@article{Paganini:2017hrr,
    author = "Paganini, Michela and de Oliveira, Luke and Nachman, Benjamin",
    title = "{Accelerating Science with Generative Adversarial Networks: An Application to 3D Particle Showers in Multilayer Calorimeters}",
    eprint = "1705.02355",
    archivePrefix = "arXiv",
    primaryClass = "hep-ex",
    doi = "10.1103/PhysRevLett.120.042003",
    journal = "Phys. Rev. Lett.",
    volume = "120",
    number = "4",
    pages = "042003",
    year = "2018"
}

@article{lipman2022flow,
    author = "Lipman, Yaron and Chen, Ricky T. Q. and Ben-Hamu, Heli and Nickel, Maximilian and Le, Matt",
    title = "{Flow matching for generative modeling}",
    eprint = "2210.02747",
    archivePrefix = "arXiv",
    primaryClass = "cs.LG",
    year = "2022"
}

@article{Albergo2022StochasticInterpolants,
  title={Building Normalizing Flows with Stochastic Interpolants},
  author={Michael S. Albergo and Eric Vanden-Eijnden},
  journal={arXiv preprint arXiv:2209.15571},
  year={2022},
  eprint={2209.15571},
  archivePrefix={arXiv},
  primaryClass={cs.LG}
}

@article{Moreno:2026mqk,
    author = "Moreno, Eric A. and Bright-Thonney, Samuel and Novak, Andrzej and Garcia, Dolores and Harris, Philip",
    title = "{AI Agents Can Already Autonomously Perform Experimental High Energy Physics}",
    eprint = "2603.20179",
    archivePrefix = "arXiv",
    primaryClass = "hep-ex",
    month = "3",
    year = "2026"
}

@article{Defranchis:2026wyw,
    author = "Defranchis, Matteo M. and Fanini, Jacopo and Fatehi, Apranik and Ganis, Gerardo and Gillin, Taj and Gouskos, Loukas and Lambrecht, Luka and Selvaggi, Michele and Stapf, Birgit",
    title = "{Modern jet flavour tagging in hadronic Z decays with archived ALEPH data}",
    eprint = "2603.06524",
    archivePrefix = "arXiv",
    primaryClass = "hep-ex",
    month = "3",
    year = "2026"
}

@article{Badea:2026klb,
    author = "Badea, Anthony and Chen, Yi and Maggi, Marcello and Lee, Yen-Jie",
    collaboration = "Electron-Positron Alliance",
    title = "{Agentic AI -- Physicist Collaboration in Experimental Particle Physics: A Proof-of-Concept Measurement with LEP Open Data}",
    eprint = "2603.05735",
    archivePrefix = "arXiv",
    primaryClass = "hep-ex",
    month = "3",
    year = "2026"
}

@article{Electron-PositronAlliance:2025fhk,
    author = "Bossi, Hannah and others",
    collaboration = "Electron-Positron Alliance",
    title = "{Energy Correlators from Partons to Hadrons: Unveiling the Dynamics of the Strong Interactions with Archival ALEPH Data}",
    eprint = "2511.00149",
    archivePrefix = "arXiv",
    primaryClass = "hep-ph",
    reportNumber = "MITP-25-057, MITHIG-MOD-24-001",
    month = "10",
    year = "2025"
}

@article{Electron-PositronAlliance:2025hze,
    author = "Badea, Anthony and others",
    collaboration = "Electron-Positron Alliance",
    title = "{Unbinned measurement of thrust in $e^+e^-$ collisions at $\sqrt{s}$ = 91.2 GeV with ALEPH archived data}",
    eprint = "2510.22038",
    archivePrefix = "arXiv",
    primaryClass = "hep-ex",
    month = "10",
    year = "2025"
}

@article{Electron-PositronAlliance:2023klx,
    author = "Chen, Yu-Chen and others",
    title = "{Long-range near-side correlation in $e^+e^-$ collisions at 183-209 GeV with ALEPH archived data}",
    eprint = "2312.05084",
    archivePrefix = "arXiv",
    primaryClass = "hep-ex",
    reportNumber = "MITHIG-MOD-23-001",
    doi = "10.1016/j.physletb.2024.138957",
    journal = "Phys. Lett. B",
    volume = "856",
    pages = "138957",
    year = "2024"
}

@article{Electron-PositronAlliance:2021kig,
    author = "Chen, Yi and others",
    title = "{Jet energy spectrum and substructure in e$^{+}$e$^{-}$ collisions at 91.2 GeV with ALEPH Archived Data}",
    eprint = "2111.09914",
    archivePrefix = "arXiv",
    primaryClass = "hep-ex",
    reportNumber = "MITHIG-MOD-21-001",
    doi = "10.1007/JHEP06(2022)008",
    journal = "JHEP",
    volume = "06",
    pages = "008",
    year = "2022"
}

@article{Electron-PositronAlliance:2019cpi,
    author = "Badea, Anthony and Baty, Austin and Chang, Paoti and Innocenti, Gian Michele and Maggi, Marcello and Mcginn, Christopher and Peters, Michael and Sheng, Tzu-An and Thaler, Jesse and Lee, Yen-Jie",
    title = "{Measurements of two-particle correlations in $e^+e^-$ collisions at 91 GeV with ALEPH archived data}",
    eprint = "1906.00489",
    archivePrefix = "arXiv",
    primaryClass = "hep-ex",
    reportNumber = "MITHIG-MOD-19-001",
    doi = "10.1103/PhysRevLett.123.212002",
    journal = "Phys. Rev. Lett.",
    volume = "123",
    number = "21",
    pages = "212002",
    year = "2019"
}

@article{ALEPH:1994ayc,
    author = "Buskulic, D. and others",
    collaboration = "ALEPH",
    title = "{Performance of the ALEPH detector at LEP}",
    reportNumber = "CERN-PPE-94-170, FSU-SCRI-95-70",
    doi = "10.1016/0168-9002(95)00138-7",
    journal = "Nucl. Instrum. Meth. A",
    volume = "360",
    pages = "481--506",
    year = "1995"
}

@techreport{Brun:1987ma,
    author = "Brun, R. and Bruyant, F. and Maire, M. and McPherson, A. C. and Zanarini, P.",
    title = "{GEANT3}",
    institution = "CERN",
    number = "CERN-DD-EE-84-1",
    url = "https://cds.cern.ch/record/1119728",
    year = "1987"
}

@article{Gaede:2022leb,
    author = "Gaede, Frank and Madlener, Thomas and Declara Fernandez, Placido and Ganis, Gerardo and Hegner, Benedikt and Helsens, Clement and Sailer, Andre and A. Stewart, Graeme and Völkl, Valentin",
    title = "{EDM4hep - a common event data model for HEP experiments}",
    doi = "10.22323/1.414.1237",
    journal = "PoS",
    volume = "ICHEP2022",
    pages = "1237",
    month = "11",
    year = "2022"
}

@article{CMS:2008xjf,
    author = "Chatrchyan, S. and others",
    collaboration = "CMS",
    title = "{The CMS Experiment at the CERN LHC}",
    doi = "10.1088/1748-0221/3/08/S08004",
    journal = "JINST",
    volume = "3",
    pages = "S08004",
    year = "2008"
}

@article{deFavereau:2013fsa,
    author = "de Favereau, J. and Delaere, C. and Demin, P. and Giammanco, A. and Lema{\^\i}tre, V. and Mertens, A. and Selvaggi, M.",
    collaboration = "DELPHES 3",
    title = "{DELPHES 3, A modular framework for fast simulation of a generic collider experiment}",
    eprint = "1307.6346",
    archivePrefix = "arXiv",
    primaryClass = "hep-ex",
    doi = "10.1007/JHEP02(2014)057",
    journal = "JHEP",
    volume = "02",
    pages = "057",
    year = "2014"
}

@article{Kobylianskii:2024sup,
    author = "Kobylianskii, Dmitrii and Soybelman, Nathalie and Kakati, Nilotpal and Dreyer, Etienne and Nachman, Benjamin and Gross, Eilam",
    title = "{Advancing set-conditional set generation: Diffusion models for fast simulation of reconstructed particles}",
    eprint = "2405.10106",
    archivePrefix = "arXiv",
    primaryClass = "hep-ex",
    doi = "10.1103/PhysRevD.110.092013",
    journal = "Phys. Rev. D",
    volume = "110",
    number = "9",
    pages = "092013",
    year = "2024"
}

@article{Dreyer:2025zhp,
    author = "Dreyer, Etienne and Gross, Eilam and Kobylianskii, Dmitrii and Mikuni, Vinicius and Nachman, Benjamin",
    title = "{Conditional deep generative models for simultaneous simulation and reconstruction of entire events}",
    eprint = "2503.19981",
    archivePrefix = "arXiv",
    primaryClass = "hep-ex",
    doi = "10.1103/14ph-482n",
    journal = "Phys. Rev. D",
    volume = "113",
    number = "3",
    pages = "032005",
    year = "2026"
}

@article{Dreyer:2024bhs,
    author = "Dreyer, Etienne and Gross, Eilam and Kobylianskii, Dmitrii and Mikuni, Vinicius and Nachman, Benjamin and Soybelman, Nathalie",
    title = "{Automated Approach to Accurate, Precise, and Fast Detector Simulation and Reconstruction}",
    eprint = "2406.01620",
    archivePrefix = "arXiv",
    primaryClass = "physics.data-an",
    doi = "10.1103/PhysRevLett.133.211902",
    journal = "Phys. Rev. Lett.",
    volume = "133",
    number = "21",
    pages = "211902",
    year = "2024"
}

@article{Hashemi:2023rgo,
    author = "Hashemi, Baran and Krause, Claudius",
    title = "{Deep generative models for detector signature simulation: A taxonomic review}",
    eprint = "2312.09597",
    archivePrefix = "arXiv",
    primaryClass = "physics.ins-det",
    reportNumber = "HEPHY-ML-23-02",
    doi = "10.1016/j.revip.2024.100092",
    journal = "Rev. Phys.",
    volume = "12",
    pages = "100092",
    year = "2024"
}

@article{Krause:2024avx,
    author = "Amram, Oz and others",
    editor = "Krause, Claudius and Faucci Giannelli, Michele and Kasieczka, Gregor and Nachman, Benjamin and Salamani, Dalila and Shih, David and Zaborowska, Anna",
    title = "{CaloChallenge 2022: a community challenge for fast calorimeter simulation}",
    eprint = "2410.21611",
    archivePrefix = "arXiv",
    primaryClass = "physics.ins-det",
    reportNumber = "HEPHY-ML-24-05, FERMILAB-PUB-24-0728-CMS, TTK-24-43",
    doi = "10.1088/1361-6633/ae1304",
    journal = "Rept. Prog. Phys.",
    volume = "88",
    number = "11",
    pages = "116201",
    year = "2025"
}

@article{GEANT4:2002zbu,
    author = "Agostinelli, S. and others",
    collaboration = "GEANT4",
    title = "{GEANT4--a simulation toolkit}",
    reportNumber = "SLAC-PUB-9350, FERMILAB-PUB-03-339, CERN-IT-2002-003",
    doi = "10.1016/S0168-9002(03)01368-8",
    journal = "Nucl. Instrum. Meth. A",
    volume = "506",
    pages = "250--303",
    year = "2003"
}

@misc{prd,
  title = {{Simultaneous PRD Submission}}
}

\end{document}